\documentclass[times,twoside,watermark]{zHenriquesLab-StyleBioRxiv}
\usepackage{url,lineno,microtype}
\usepackage[onehalfspacing]{setspace}
\usepackage[english]{babel}
\usepackage{soul}
\usepackage{xcolor}
\usepackage{ulem}

\leadauthor{Woringer} 

\begin{document}

\title{Anomalous subdiffusion in living cells: bridging the gap between experiments and realistic models through collaborative challenges.}
\shorttitle{A challenge for anomalous subdiffusion in living cells}

\author[1,2,3,7]{Maxime Woringer}
\author[4,7]{Ignacio Izeddin}
\author[5,7,\Letter]{Cyril Favard}
\author[4,7,\Letter]{Hugues Berry}

\affil[1]{Unité Imagerie et Mod\'elisation, CNRS UMR 3691, and C3BI (Center of Bioinformatics, Biostatistics and Integrative Biology), CNRS USR 3756, Institut Pasteur, 75015 Paris, France}
\affil[2]{Sorbonne Universit\'es, CNRS, 75005 Paris, France}
\affil[3]{Department of Molecular and Cell Biology, Li Ka Shing Center for Biomedical and Health Sciences, and CIRM Center of Excellence in Stem Cell Genomics, University of California, Berkeley, California 94720, USA}
\affil[4]{Institut Langevin, ESPCI Paris, CNRS, PSL University, 1 rue Jussieu, Paris 75005, France}
\affil[5]{Membrane Domains and Viral Assembly, Institut de Recherche en Infectiologie de Montpellier, CNRS UMR 9004, Montpellier, France}
\affil[6]{Inria, Lyon, F-69603, Villeurbanne, France, and Universite de Lyon, LIRIS UMR5205, F-69621, Villeurbanne, France}
\affil[7]{GDR Imabio, CNRS, Lille, France}

\maketitle

\begin{abstract}
The life of a cell is governed by highly dynamical microscopic processes. Two notable examples are the diffusion of membrane receptors and the kinetics of transcription factors governing the rates of gene expression. Different fluorescence imaging techniques have emerged to study molecular dynamics. Among them, fluorescence correlation spectroscopy (FCS) and single-particle tracking (SPT) have proven to be instrumental to our understanding of cell dynamics and function.
The analysis of SPT and FCS is an ongoing effort, and despite decades of work, much progress remains to be done. In this paper, we give a quick overview of the existing techniques used to analyze anomalous diffusion in cells and propose a collaborative challenge to foster the development of state-of-the-art analysis algorithms. We propose to provide labelled (training) and unlabelled (evaluation) simulated data to competitors all over the world in an open and fair challenge. The goal is to offer unified data benchmarks based on biologically-relevant metrics in order to compare the diffusion analysis software available for the community.
\end {abstract}


\begin{corrauthor}
cyril.favard\at irim.cnrs.fr, hugues.berry\at inria.fr
\end{corrauthor}

\section*{Introduction}
The life of a cell is governed by highly dynamical microscopic processes occurring at different space and time scales from single macromolecules up to organelles. Optical microscopy provided four decades ago the first measurements of biomolecule motion in cells. First by fluorescence recovery after photobleaching (FRAP) \cite{axelrod_mobility_1976} and fluorescence correlation spectroscopy (FCS) \cite{magde_thermodynamic_1972}, and more recently with the help of single particle tracking (SPT) \cite{Geerts1987,Geerts1991}. 
Several factors have colluded to popularize these techniques in many biophysics and biology labs: i) the development of highly sensitive detectors, ii) the emergence of genetically encoded fluorescent protein labelling in the late 90s \cite{Heim12501,HEIM1996178,Matz1999}, and iii) the advent in the years 2000-2010 of far-field super-resolution microscopy\cite{Hell1994,Betzig2006b,hess_ultra-high_2006,Klar8206,rust_sub-diffraction-limit_2006}. All these technological efforts have granted us access to the monitoring of molecular motion in cells with unprecedented spatial (down to single molecule) and temporal resolution \cite{manley_high-density_2008,eggeling_direct_2009}. The adoption of these techniques has been paramount in the advancement of the understanding of cell organisation and dynamics \cite{Chattopad_2019,Liu_rev_2019,Priest1117}.

While acquiring sufficient experimental data sets used to be a limiting factor, these technological advances combined with data acquisition parallelization provide nowadays huge amounts of data available for analysis of molecular motion inside the cell. In turn, the richness of this data has unravelled an unforeseen complexity and diversity of mechanisms for biomolecule motion in cells. Therefore, many efforts are devoted to analyze data provided by FCS or SPT with direct or inference approaches.

However, choosing the appropriate algorithms to analyse the complexity of the observed phenomena is still an important challenge. Indeed, the richness of experimental data often makes it difficult to determine which are the physical models to be considered and which are the relevant biophysical parameters to be estimated from them. We review and address this issue in this perspective.

We will first briefly review key anomalous diffusion models relevant to cell biology and summarily describe some of the existing techniques to either infer model parameters or to perform model selection. We will discuss the relevance of numerical simulations and the importance of designing realistic data sets closely mimicking the results obtained in experiments on biological samples. We will also highlight the often overlooked limitations in current acquisition methods and emphasize the role of experimental noise and biases of the aforementioned techniques. Finally, we will present and advocate in favour of the development of comprehensive sets of simulated data and metrics, allowing the community to objectively evaluate existing and new analysis tools. Our hope is that this work will instigate an open discussion about the limitations and challenges of analysing and modelling diffusion of molecules in the complex environment of the cell.

\section*{Brownian vs anomalous diffusion}

Maybe one of the best-known result of the theory of Brownian diffusion is that the mean squared displacement (MSD) of a random walker scales linearly with time, and is proportional to the diffusion coefficient of the fluid in which diffusion takes place. With $x(t)$ being the position of the random walker at time $t$ (in one dimension), this means that the MSD $\left\langle {x(t)}^2\right\rangle = 2 D t$, where $\langle \cdot \rangle$ denotes ensemble averaging and $x(0)=0$. However, Brownian diffusion does not explain the physics of disordered systems. Interestingly, an ubiquitous observation in cell biology is that the diffusive motion of macromolecules and organelles is anomalous, i.e. the MSD change with time is typically characterized by a sublinear increase. In most instances, this sublinear increase of the MSD with time can be fitted to a power-law relation $\left\langle {x(t)}^2\right\rangle \propto t^\alpha$ with exponent $\alpha<1$, which justifies the vocable of ``subdiffusion''. Subdiffusion is usually attributed to cellular crowding, spatial heterogeneity or molecular interactions.
Another possibility of anomalous diffusion is superdiffusion, with $1<\alpha<2$. Indeed a lot of process in biology exhibit active transport or combinations of active and random motions.

Anomalous diffusion in cells is therefore a very active area of research involving biophysics, cell biology, statistical physics and mathematical modelling.

When confronted to a set of data retrieved from FCS or SPT experiments, the first question that one needs to answer is whether the measured \sout{sub}diffusion is indeed a manifestation of an anomalous process. Often, a combination of several normal diffusion mechanisms or experimental artefacts gives rise to an apparent \sout{sub}diffusion. If an anomalous diffusion --characterized by a power law scaling of the MSD with time-- can be identified, establishing the physical model behind the diffusion process can shed light on the molecular mechanisms driving the motion of the molecule of interest.

Below, we will first focus on three classical models for anomalous subdiffusion and their common biological interpretation, namely the continuous-time random walk (CTRW) model, the fractional Brownian motion (fBm) model, and random walks on fractal and disordered systems (for a review, see, e.g. \cite{hoefling_anomalous_2013}), then we will briefly describe different models covering super-diffusion processes that can be encountered in cells such as run and tumble model, Levy flights and super-diffusive fBm.

The \textbf{continuous-time random walk} model is a generalization of a random walk in which the diffusing particle waits for a random time between jumps. More generally, when the distribution $\phi(\tau)$ of waiting times $\tau$ is long-tailed and cannot be averaged (with e.g. $\phi(\tau)\propto \tau^{-(1+\alpha)}$ and $0<\alpha<1$ ), the ensemble-averaged MSD shows anomalous scaling with a power law. A straightforward interpretation of a CTRW in the context of molecular biology is assimilating the waiting times to interactions of the molecule with an immobile substrate (at the relevant temporal and spatial scales). It is important to note that an interaction with a characteristic residence time does not fulfill the conditions of the model. Interestingly, however, the waiting-time distribution of non-specific interactions, abundant in the cell, might be non-averageable and thus CTRW a good microscopic model for one type of anomalous subdiffusion in the cell. It has been proposed to govern the cytosolic diffusion of nanosized objects in mammalian cells \cite{Etoc2018} and it has also been used to explain the lateral motion of potassium channels in the plasma membrane of cells \cite{Wiegel2011}.

The \textbf{fractional Brownian motion} model is a different generalization of Brownian diffusion in which the jumps between lag times follow a normal distribution but respect a correlation function given by $\left\langle x(t)x(s)\right\rangle=1/2(t^{2H}+s^{2H}-(t-s)^{2H})$ for $t>s>0$. A fBm process is thus characterized by the Hurst index $H$, ranging between $0$ and $1$. The value of $H$ determines the type of jump dependence in the fBm process, such that $H > 1/2$ indicates a positive correlation between the increments, Brownian motion is achieved for $H = 1/2$, and the increments are negatively correlated when $H < 1/2$. The MSD of a fBm is given by $\left\langle {x(t)}^2\right\rangle \propto t^{2H}$, which, again, encompasses Brownian diffusion for $H=1/2$ and yields subdiffusion for $H<1/2$ or superdiffusion for $H>1/2$ (see below). The fBm model describes faithfully the diffusion of particles in a viscoelastic fluid \cite{ernst_fractional_2012}, and it has been often argued that molecular crowding in the cell gives rise to microviscosity and therefore to anomalous diffusion. It was proposed as the model of telomere diffusion in nucleus \cite{Kepten2011,burnecki_universal_2012}.

Another possible model for anomalous diffusion in the cell is that of \textbf{random walks on fractal media and disordered systems}. Fractals are self-similar mathematical objects built upon the repetition of simple rules and characterized by a non-integer number: the \textit{fractal dimension}. Although still under debate, some authors have proposed that chromatin organization follows, as a first order approximation, a fractal structure, and estimates of its fractal dimension have been proposed~\cite{recamier_single_2014}. Random walks on fractals are subdiffusive due to the spatial correlation of displacements, and the power law scaling factor of the MSD with time is given by $2/d_w$, where $d_w$ is the \textit{dimension of the walk} that is specific to the fractal. Although the pertinence of a fractal network model to describe molecular diffusion is still up to debate, it is justified to attempt to integrate the multiscale characteristics of the cell organization to such fractal model.

Amongst the existing superdiffusive motion in cells is the
\textbf{run-and-tumble process}, which consists of alternating phases of fast active and slow passive motion leading to transient anomalous diffusion \cite{Shebani_run_tumble_2019}. Initially observed for bacteria motion it has recently been used to describe molecular motions in cells such as the motion of motors along cytoskeletal filaments. Motor proteins perform a number of steps (run) until they randomly unbind from the filaments and diffuse in the crowded cytoplasm (tumble) before rebinding \cite{Hafner2016}. The same could also stand for transcription factors in the nucleus searching for their initiation codon, alternating successively diffusion and 1D sliding along the DNA.
\textbf{Superdiffusive fBm} which is characterized by an Hurst index $H>1/2$ has been described as the intracellular motion of particles in the super-crowded cytoplasm of a amibae \cite{Reverey2015}. Finally, \textbf{Levy flights}, has previously been proposed for intracellular actin-based transport mediated by molecular motors \cite{Bruno2009} and recently in the case of a membrane targeting C2 protein \cite{Campagnola2015}.

Note that by no means the above described models exhaustively cover the range of models that are known to exhibit anomalous diffusion (see e.g. \cite{metzler_anomalous_2014,Pavlos_2014,Lenzi_2016}). However the CTRW, fBM, and random walks in a fractal models have been extensively studied; more importantly, they have the potential to map parameters of the model to relevant biological and biophysical features. Therefore, we will limit our discussion to the aforementioned cases, and how they can be used to analyse and interpret experimental data obtained by FCS and SPT.


\section*{Which methods to correctly analyse diffusive process?}
\subsection*{Fluorescence Correlation Spectroscopy}
The principle of FCS consists in measuring the temporal variations of molecular concentration at a given position within the volume of a biological sample. This is achieved by monitoring the temporal fluctuations of fluorescence signal emitted by the molecules present in the observation volume, which is excited with a focused laser. The underlying assumption of FCS is that the system is in a dynamic equilibrium and therefore the signal fluctuation can be correlated to the diffusion of molecules within the observation volume. While the amplitude of the fluctuations relates to the number of molecules in the observation volume, the decay of their autocorrelation in time depends on their mobility. 

A  typical  FCS  set-up  consists  of  an  illumination  laser and  a  confocal  microscope  with  a  fast single-channel single-photon  detector. The laser beam illuminates the detection volume with, usually, a Gaussian intensity profile and excites the fluorophores in the focal volume. The  emitted  fluorescent  light  is  collected  by  the  detector and it  depends  on  the  fluctuations of the local concentration of the labelled  molecules.

\begin{figure*}[h!]
\begin{center}
\includegraphics[width=18cm]{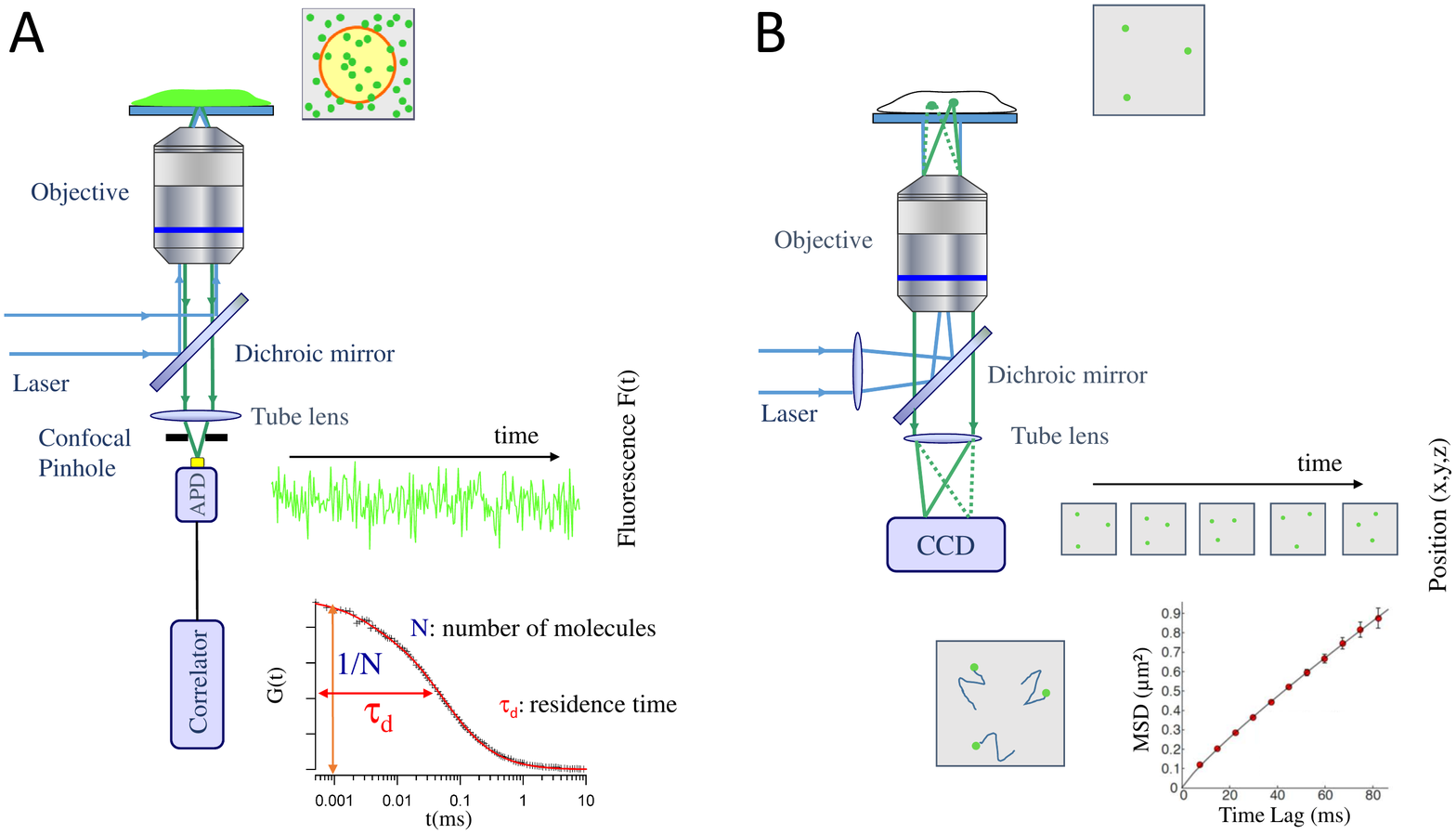}
\end{center}
\caption{\textbf{Figure 1. Schematic view of the typical setup used in fluorescence correlation spectroscopy (A) and single/multiple particle tracking (B) experiments.}
\textbf{A}: A laser is focused on the fluorescently labelled sample by the objective of a microscope. The fluorescence is then collected by the objective and focused in a confocal way (using a pinhole) on a single photon counting detector (avalanche photodiode, APD). This detector records the fluctuation of fluorescence emission within the confocal volume of the sample. A direct link to an electronic correlator authorize on line generation of the autocorrelogram.
\textbf{B}: A laser is focused at the back focal plane of a microscope objective in order to obtain a full field illumination of the sample. The fluorescence emitted by each single particle present in the illumination field is then directly imaged on a sensitive camera (Charge Coupled Device, CCD). A movie is obtained and the post processing of this movie allow tracking of the individual emitter and latter on, generation of Mean Square Displacement (MSD) as a function of lag time curves.}\label{fig:1}
\end{figure*}

Parameters such as the average number of molecules (N) and their mean residence time ($\tau_d$) in the confocal volume (surface) can be obtained either directly from this fluorescence intensity fluctuation measurement or indirectly by a temporal auto-correlation of this fluctuation. The second method is the most popular approach for FCS data analysis (see Fig. \ref{fig:1}). The main drawback of standard FCS is the lack in directly monitoring possible spatial and/or temporal heterogeneities that will give rise to deviation from pure Brownian motion. Several approaches have been proposed to overcome this issue including spot variation FCS (sv-FCS) \cite{wawrezinieck_fluorescence_2005,eggeling_direct_2009}, line scanning FCS and STED-FCS \cite{PETRASEK20081437,honigmann_scanning_2014} 
, as well as imaging approaches such as (spatio)-temporal imaging correlation spectroscopy ((S)TICS), raster imaging correlation spectroscopy (RICS) \cite{digman_measuring_2005} or more recently whole plane Imaging FCS (Im-FCS)\cite{Kannan2006}.
With the development of commercial microscopes coupled to FCS capabilities, this technique and its derivatives are now becoming more and more popular in biology labs.

A large range of dynamic processes leading to concentration fluctuations (i.e, diffusion, flow, chemical reactions and different combinations of these) has been investigated to generate corresponding analytical expressions of the temporal autocorrelation curve $G(t)$ in the case of Gaussian (laser confocal) illumination/detection geometry (for a review, see \cite{Elson2011} and references therein). For instance, in the case of a Brownian motion in 2D, $G(t)=1/\{\bar{N}(1+4Dt/w^2)\}$ where $w$ is the size of the beam waist and $\bar{N}$ is the average number of molecules in the observation volume.
The main approach to diffusive process identification and quantification in FCS consists in non linear least square fitting of experimental autocorrelation curves using above described analytical expressions and discriminate amongst these models which suits the best using various statistical test. Although it can deliver quantitative values of the parameters of the statistically chosen model of motion, it could be strongly biased, in particular for complex motions. A Bayesian approach to single spot FCS correlogram analysis has been proposed to discriminate between different models without bias \cite{He2012,guo_bayesian_2012} 

Another way to discriminate between different types of motion is to explore space and time with FCS using svFCS for example. svFCS offers the opportunity to generate so-called "diffusion-laws" by plotting changes in the residence time ($\tau_d$) as a function of the surface (i.e. laser waist) explored $w^2$. This has enabled to directly identify deviations from pure Brownian motion in the plasma membrane of cells \cite{lenne_dynamic_2006} or anomalous diffusion occurring, either during first order lipid phase transition \cite{favard_fcs_2011} or in non-homogeneous fluids, gels and crowded solutions \cite{Banks_anomalous_FCS,Masuda_2005}.
It has been recently extended to the line-scanning STED-FCS \cite{Schneider2018} and to Im-FCS \cite{Veerapathiran2018}.

\subsection*{Single/Multiple Particle(s) Tracking}
While the concentration of the subset of fluorescent molecules within a confocal volume in FCS experiments is close to the single-molecule regime, the measurement gauges the average motion of the ensemble of molecules diffusing in and out the observation spot. Conversely, SPT is by construction a single-molecule approach, monitoring thus the motion of individual molecules. One of the strengths of SPT is the potential to capture rare events or behaviours that would otherwise be buried within an average.

The principle of SPT experiments is simple, it consists in retrieving the changes in position of individual molecules within the sample of interest, i.e. the time series of two-dimensional or three-dimensional coordinates of the molecule location. This is achieved in two stages: firstly by estimating the centroid of the measured point spread function (PSF) of each detected individual emitter, and secondly by linking the trajectory of the same molecule between consecutive images. Importantly, the accuracy at which one is able to pinpoint the molecule position depends only on the signal-to-noise ratio of the measured PSF, obtaining sub-wavelength accuracy typically in the order of $\sim$10 nm.

The basic SPT experimental setup consists of an excitation laser, a high NA objective, a set of dichroic and filters to separate the excitation and emission wavelengths, a tube lens, and a highly sensitive camera capable of detecting single fluorophores (see Fig. \ref{fig:1}). The laser is focused on the back focal plane of the objective to obtain a wide-field illumination configuration, which can be adjusted to total internal reflection (TIRF) or highly inclined illumination (HILO)~\cite{Tokunaga2008} to increase the SNR when studying molecular dynamics in cellular membranes or at the interior of cells, respectively. The fluorescence light is collected by the same objective, and an image of the single emitters is formed on the camera plane via the tube lens~\cite{manley_high-density_2008,izeddin_single-molecule_2014}.

The amount of retrieved information about the biological system from an SPT assay depends on the nature of the experiment. The study of a slowly diffusing transmembrane protein will yield much longer traces than a fast diffusing transcription factor in the nucleus. In the latter case, the traces will be limited to the number of images in which the tracked particle remains within the depth of focus around the image plane, unlike the former case where photobleaching is the limiting factor.

The classical analysis of a set of trajectories consists in computing the dependence of the MSD (time-average or ensemble-average) over time from the distribution of jumps at increasing lag times defined by the camera acquisition, typically in the order of tens of ms. However, as we will see in the following section, different approaches and estimators have been proposed in order to analyze and interpret SPT data to its full extent.
In comparison to FCS, the analysis of SPT has been intensively investigated, and one can distinguish several families of techniques (see also for reviews: \cite{metzler_analysis_2009,recamier_single_2014,ernst_probing_2014}). In the field of stochastic processes, the inference of a diffusion coefficient from a sampled process is a common problem (see for instance \cite{florens-zmirou_estimating_1993,hoffmann_estimating_2001}). However, this theory cannot be applied when moving to experimental trajectories, and other approaches have been proposed.

\subsubsection*{MSD-based techniques}
A first family of SPT analysis algorithms tries to perform robust MSD inference. The use of MSD to study diffusion was introduced by Einstein in 1906, and was revived in biology by \citep{qian_single_1991}. MSD analysis can either be performed by inferring a diffusion coefficient from a single trajectory (a setting studied in \cite{michalet_optimal_2012}) or by pooling various trajectories \cite{liu_2014}, and many refinements and estimators based on the MSD have been proposed \cite{michalet_mean_2010,boyer_optimal_2012}. 

When inferring kinetic parameters from a series of single trajectories, one faces the issue that for common trajectory lengths obtained in nuclear SPT (length of $<<20$ points per track) and common localization error, inaccuracy might reach 100\% \cite{michalet_optimal_2012,hansen_robust_2018}. As such, any approach that uses MSD on short trajectories should be evaluated with great care. For longer trajectories (such as diffusion in a membrane), approaches have been proposed that can segment trajectories based on the type of motion \cite{monnier_bayesian_2012}.

\subsubsection*{Hidden Markov Models (HMMs)}
A second family of SPT analysis algorithms derives from Markov models and Hidden Markov Models. Most of them were derived to perform trajectory segment classification, the hidden variable inferred being the state of diffusion, or the current diffusion coefficient. For instance, \cite{monnier_inferring_2015} introduces the HMM-Bayes technique to infer whether a trajectory segment is in one (or several) diffusive or active transport states. Moreover, \cite{slator_detection_2015} implemented the inference of localization noise to infer switches in diffusion coefficient within one trajectory. A similar approach was used to detect confinement \cite{slator_hidden_2018}.

These methods often rely on a fixed number of states, which comes from significant mathematical limitations. Some of these limitations were overcome using so-called variational Bayesian inference \cite{blei_variational_2016}. The prototypical algorithm performing variational Bayesian inference on a HMM is vbSPT \cite{persson_extracting_2013}. This algorithm can estimate the number of diffusive states and progressively consolidate increasing information about these states as trajectories are analyzed. The algorithm was further refined to incorporate the estimate of localization error \cite{linden_variational_2018-1}.

\subsubsection*{Inferring maps of diffusion coefficients}
A third family of SPT analysis algorithms not only infers the diffusion coefficient over the population of diffusing molecules, but also a spatial map of diffusivity \cite{masson_inferring_2009,el_beheiry_inferencemap_2015}. This approach has been pioneered in membranes, where a high density of tracks can easily be obtained. An extension of this approach using an overdamped Langevin equation of the single molecule motion has shed new lights on HIV-1 assembly within living cells \cite{floderer_single_2018}. These promising techniques have not been tested beyond membrane molecules, but the high diffusion coefficients of freely diffusing cellular proteins might render such a map difficult to establish. Moreover, unlike in membranes, proteins can reside at the same location with different diffusion coefficients, depending on whether they are interacting with a given structure or not.

\subsubsection*{Inferring anomalous diffusion}
Many approaches have been proposed to infer anomalous diffusion in cells; some of them are reviewed in \cite{guigas_sampling_2008}. A direct technique can be used by fitting the MSD with a power law to estimate the anomalous diffusion coefficient $\alpha$. However, alternative techniques have been proposed, many of them focused on the inference of model-specific parameters, or on techniques to distinguish between types of anomalous diffusion.

Several methods have been proposed to infer diffusion parameters for several anomalous diffusion models. For the case of diffusion in disordered (fractal) media, \cite{shkilev_kinetic_2018} proposes estimators that can be applied to SPT, FCS and FRAP. For the case of fractional Brownian motion, techniques to infer both the anomalous diffusion coefficient ($\alpha$) and the generalized diffusion coefficient ($D_\alpha$) have been proposed. The former approach \cite{krog_bayesian_2018} takes into account noise (localization error) and drift, and uses Bayesian inference. The latter \cite{boyer_ergodic_2013} relies on squared displacements and uses least squares to estimate $D_\alpha$.

Conversely, instead of trying to estimate the parameters of a known model, a key question is to distinguish between various anomalous diffusion models. A prototypical approach \cite{robson_inferring_2012} used Bayesian inference to distinguish between Brownian, anomalous, confined and directed diffusion, and uses the propagators associated with each different diffusion model. However, \cite{hellmann_challenges_2011} found using simulations that it is very hard to distinguish between fBm and diffusion on a fractal when localization noise is present, both in SPT and FCS. The authors used a combination of techniques for the inference, including MSD and $p$-variation techniques. In \cite{burnecki_universal_2012}, the authors propose a series of tests to "unambiguously" identify fBm, by progressively proving that several other models are wrong. Other tests were proposed to distinguish fBm from a CTRW using a test based on $p$-variations \cite{magdziarz_fractional_2009}. The $p$-variations are the finite sum of the $p$-th powers of the increments of the trajectory. Finally, approaches inferring the mean first passage time of a particle were used to distinguish between CTRW and diffusion in fractals \cite{condamin_first-passage_2007,condamin_probing_2008}.

Many other families of techniques to identify types of diffusion have been proposed. Some relied on maximum likelihood estimates \cite{thapa_bayesian_2018}, auto-correlation functions \cite{weber_analytical_2012} or on more exotic estimators \cite{vestergaard_optimal_2014}. Another line of progress was made in the type of models being simulated. For instance, \cite{amitai_chromatin_2018} introduced a model in which TFs can bind and rebind in a dense chromatin mesh. This model was successively fitted to explain anomalous diffusion of CTCF dynamics \cite{hansen_guided_2018}.

Finally, we note that many models were developed to infer trapping potential in membranes (\cite{turkcan_bayesian_2012,masson_mapping_2014} for instance). We do not review them here since their application seems limited to membranes.

\subsection*{Strengths \& limitations of the two techniques}
A strong limitation is that the experimental context, either in FCS or in SPT, may lead to spurious determination of anomalous diffusion. In other words, specific experimental parameters (low statistics, location noise, spatial confinement, etc.) and/or inappropriate anaysis of the data can lead to incorrectly conclude that the diffusion exponent $\alpha \neq 1$. Those artifacts concern both SPT~\cite{Martin_2002} and FCS~\cite{Banks_2016}. This is for instance the case if $\alpha$ is determined by a fit of the MSD or the autocorrelation with time and the statistical power is low (low sampling of the time points or short trajectories in SPT, low signal/noise at small or large times in FCS). To avoid such caveats, model selection must use more elaborate approaches to unambiguously demonstrate and characterize an underlying complex diffusion process.

So far, most of the inference tools available in the literature only partially account for the biases detailed above, and are usually limited in terms of the anomalous diffusion models they consider. For instance, in~\cite{hansen_robust_2018}, the authors showed that an algorithm not taking into account localization error was likely to improperly estimate diffusion coefficients. Similarly, the fact that the observed proteins diffuse in a confined volume leads to a sublinear MSD, a phenomenon that has been widely documented and that needs to be taken into account to properly distinguish between genuine anomalous diffusion and mere confinement effect. Similarly, tracking errors (misconnections between tracks) can also look like anomalous diffusion.

Some of these biases can be minimized at the acquisition step (for instance by using fast frame rates and low labeling density~\cite{hansen_robust_2018}), other need to be explicitly taken into account in the model. As of today, most inference algorithms available have not been benchmarked against realistic imaging conditions. Furthermore, a general realistic inference algorithm is still missing.

\section*{Conclusion: the need for controlled benchmarks}
    Confronted with the variety of approaches described above, one would like to know the performance of each approach on typical representative datasets. For the comparison to be fair, this demands two main ingredients: (\emph{i}) the existence of a reference dataset, or benchmark -- possibly one reference dataset for each main classes of experimental methods and (\emph{ii}) a fair, objective, transparent and open comparison process, with datasets, comparison procedures and performance results that are clearly stated and publicly available. Several fields in computer science have been using open community competitions to organize the process and produce open benchmarks for the community. Computer vision, applied machine learning or time series forecasting, among many others, have a long tradition of leveraging these competitions. The strategy has been widely successful because it parallelizes research along a vast community of high-skilled researchers. Internet platforms or services are even available to that purpose, including, among many others, Kaggle (~\url{www.kaggle.com}) or DrivenData (~\url{www.drivendata.org}). This increases further the size of the competing community, and the richness of the proposals. In fact, in addition to providing reference datasets and benchmarks, open competitive challenges can also foster the emergence of radically new approaches to the open problem at hand. Many of these competitive challenges are concerned with biomedical applications (for instance, \url{http://dreamchallenges.org} or~\url{https://grand-challenge.org}), including several revolving around microscopy (see e.g.~\url{https://cremi.org}). Recently, a series of consecutive community competitions for single-molecule imaging have involved dozens of labs and focused on tracking algorithms~\cite{chenouard_objective_2014}, and 2D and 3D localization for super-resolution~\cite{Sage2019}. Finally, another challenge has also been set up recently to infer the anomalous diffusion exponent from particle trajectories \url(https://competitions.codalab.org/competitions/23601). 

In practice, an important feature of competitive challenges is to provide labelled data examples that the participants will be able to use as a training set. Indeed according to standard machine learning practice, this training dataset must be distinct from the test set, that includes the data used to estimate the performance of the algorithm. The organizers therefore usually publish two datasets (training dataset and test), of which only the training dataset comes with the label of each examples -- only the organizers know the true label of the test dataset. After training, the results of the challenge is based on some quantification of the performance of the participant tools on the test set, although performance on the learning set can also be communicated as a way to judge overtraining/generalization capacities. In many cases however, it is not possible to provide the ``true'' label of experimental data, because such a gold standard does not exist. In this case, computer simulations can be used to generate synthetic data, as long as these simulations are realistic enough that the performance of the algorithms is not different than their performance on real experimental measurements. In the recent challenges on super-resolution, training and test data were a combination of computer-generated data and experimental data. Computer-generated data gives a clear access to ground truth whereas experimental data incorporate uncharacterized biases that can affect the inference process.

Here we propose to organize an international open collaborative challenge for the quantification and analysis of molecule movements in living cells via SPT and FCS. To date, the generation of realistic computer-simulated data has been hampered by the number of experimental biases to be taken into account, and by the diversity of the diffusion models, in particular for anomalous diffusion. For the challenge, we will generate both SPT and FCS data from the same set of simulated trajectories and in different modalities (2D in membranes and 3D in the nucleus) using a dedicated open source simulation software, simSPT (\url{https://gitlab.com/tjian-darzacq-lab/simSPT}), that is freely available to the participants to generate their own additional training sets if needed.

The challenge will be organized around various sub-challenges that represent the main classes of experimental situations (high-density short trajectories in membranes, less dense long trajectories in membranes, very short trajectories in the nucleus) and the main types of Brownian and anomalous diffusion (Brownian motion, fractional Brownian motion, continuous-time random walks and diffusion on fractals), and mixtures thereof. In the long run, we will also propose sub-challenges where the molecule dynamics depends on the location, to emulate localized spatial heterogeneity in the dynamics (local potentials, position-dependent diffusion coefficients). Moreover, we will progressively propose two challenge categories. In parameter inference challenges, the models used to generate the trajectories (Brownian motion, anomalous diffusion, ...) will be given and the task will be to infer as precisely as possible the value of the parameters used for the generation. In model selection challenges, the goal will be to infer what model was used to generate the data given a known limited list of models.

Finally, we are aware that it may well be that no generic tool is able to solve all the sub-challenges evoked above. We are also aware that the difficulty of each sub-challenges can be quite variable. We therefore propose to start with the simple challenges and work in collaboration with the community involved in the analysis of molecular dynamics in living cells, to progressively climb the steps toward the more difficult sub-challenges. In this strategy, maintaining an open communication channel between the organizers and the participants is paramount. To this aim, we propose to start with a mailing list that will be used to support this communication. Every interested individual is therefore welcome to subscribe to the mailing list of the challenge by visiting \url{https://listes.services.cnrs.fr/wws/info/diffusion.challenge}. Once registered in the mailing list through this website, participants will be able to exchange with the organizers and they will receive the instructions to access the datasets of the challenge.


\section*{Conflict of Interest Statement}

The authors declare that the research was conducted in the absence of any commercial or financial relationships that could be construed as a potential conflict of interest.

\section*{Author Contributions}
MW, II, CF and HB developed these perspectives and wrote the manuscript.

\section*{Funding}
This work was partly funded by the CNRS-supported GDR ImaBio, \url{http://imabio-cnrs.fr}.


\bibliographystyle{frontiersinHLTH&FPHY} 
\section*{Bibliography}
\bibliography{Biblio5}

\end{document}